\documentclass[useAMS,usenatbib]{mn2e}
\usepackage{graphicx}
\usepackage{amsmath}
\usepackage{upgreek}

\newcommand{\ha}{H$\alpha$} 
\newcommand{\hbeta}{H$\beta$}

\newcommand{\helium}{He~{\sc i}}
\newcommand{\heliumb}{He~{\sc ii}}

\newcommand{\nitrogen}{[N~{\sc ii}]}
\newcommand{\nitrogena}{[N~{\sc i}]}
\newcommand{\oxygeniii}{[O~{\sc iii}]}
\newcommand{\oxygeni}{[O~{\sc i}]}

\newcommand{\sulfurt}{[S~{\sc ii}]}

\newcommand{\degree}{$^{\circ}$}
\def\vhel{\ifmmode{V_{{\rm HEL}}}\else{$V_{{\rm HEL}}$}\fi}
\def\vsys{\ifmmode{V_{\rm sys}}\else{$V_{\rm sys}$}\fi}
\def\kms{\ifmmode{~{\rm km\,s}^{-1}}\else{~km~s$^{-1}$}\fi}
\def\vlsr{\ifmmode{v_{\rm lsr}}\else{$v_{\rm lsr}$}\fi}

\title[The bipolar planetary nebula Abell~14]
{Deciphering the bipolar planetary nebula Abell~14 with 3D ionization and morphological studies}

\author[S. Akras et al.] 
{S.~Akras$^{1}\thanks{e-mail:akras@astro.ufrj.br}$, N.~Clyne$^{1,2}$, P.~Boumis$^{3}$, H.~Monteiro$^4$, D.~R.~Gon\c calves$^1$, M.~P.~Redman$^2$, 
\newauthor
S.~Williams$^3$\\
$^1$Observat\'orio do Valongo, Universidade Federal do Rio de Janeiro, Ladeira Pedro Antonio 43 20080-090 Rio de Janeiro, Brazil\\
$^2$Centre for Astronomy, School of Physics, National University of Ireland Galway, University Road, Galway, Ireland\\
$^3$Institute for Astronomy, Astrophysics, Space Applications and Remote Sensing, National Observatory of Athens,\\
I. Metaxa \& V. Pavlou, Penteli, GR--15236, Athens, Greece\\
$^4$Instituto de F\'isica e Qu\'imica, Universidade Federal de Itajub\'a, Brasil}
\begin{document}  

\date{Received **insert**; Accepted **insert**}

\pagerange{\pageref{firstpage}--\pageref{lastpage}}

\maketitle
\label{firstpage}

\begin{abstract}
Abell~14 is a poorly studied object despite being considered a born again planetary nebula.
We performed a detailed study of its 3D morphology and ionization structure using the \textsc{shape} and \textsc{mocassin} codes.
We found that Abell~14 is a highly evolved, bipolar nebula with a kinematical age of $\sim$19,400~yr for a distance of 4~kpc. 
The high He abundance, and N/O ratio indicate a progenitor of 5~$M_{\odot}$ that has experienced the third dredge--up and hot bottom 
burning phases. The stellar parameters of the central source reveal a star at a highly evolved stage near to the white 
dwarf cooling track, being inconsistent with the born again scenario. The nebula shows unexpectedly strong \nitrogena\ $\lambda 
5200$ and \oxygeni\ $\lambda 6300$ emission lines indicating possible shock interactions. Abell~14 appears to be a member of a small group of highly evolved, 
extreme Type--I PNe. The members of this group lie at the lower-left corner of the PNe regime on the \nitrogen/\ha\ vs. \sulfurt/\ha\ diagnostic 
diagram, where shock--excited regions/objects are also placed. The low luminosity of their central stars, in conjunction with the large physical 
size of the nebulae, result in a very low photo--ionization rate, which can 
make any contribution of shock interaction easily perceptible, even for small velocities.

\end{abstract}


\section{Introduction}

Planetary nebulae (PNe) represent the final product of the interaction between a slow and dense stellar 
wind that originates from an Asymptotic Giant Branch (AGB) star, and a subsequent, faster and 
tenuous wind expelled between the very late AGB phase and at the beginning of the post--AGB phase. 
The interaction between these two stellar winds results in the formation of a PN (Interacting stellar wind model; 
Kwok et al. 1978). This model can adequately explain the formation of spherically symmetric PNe.  

Nevertheless, the majority of PNe and proto--PNe show asymmetric and complex morphologies 
(Boumis et al. 2003, 2006; Parker et al. 2006; Manchado et al. 2011; Sahai et al. 2011;  Sabin et al. 2014) that can 
not be explained by the ISW model. As a consequence, the development of a more general picture, which employs an 
axisymmetric equatorial density enhancement in the AGB wind (generalized ISW; see Balick \& Frank 2002), 
has been established as the standard model to explain the formation of non--spherically symmetric PNe. 


\begin{figure*}
\centering
\includegraphics[scale=0.28]{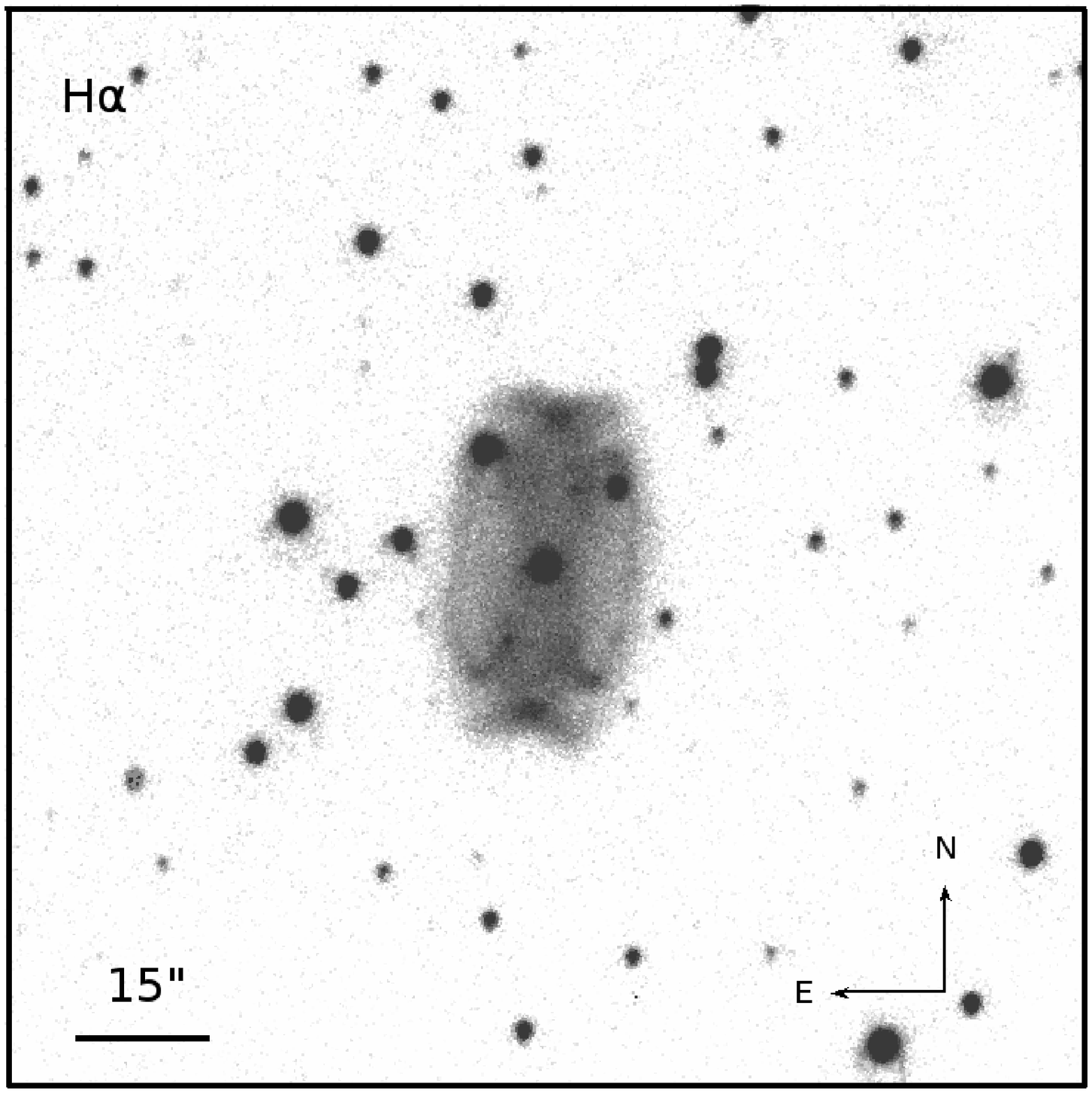}
\includegraphics[scale=0.28]{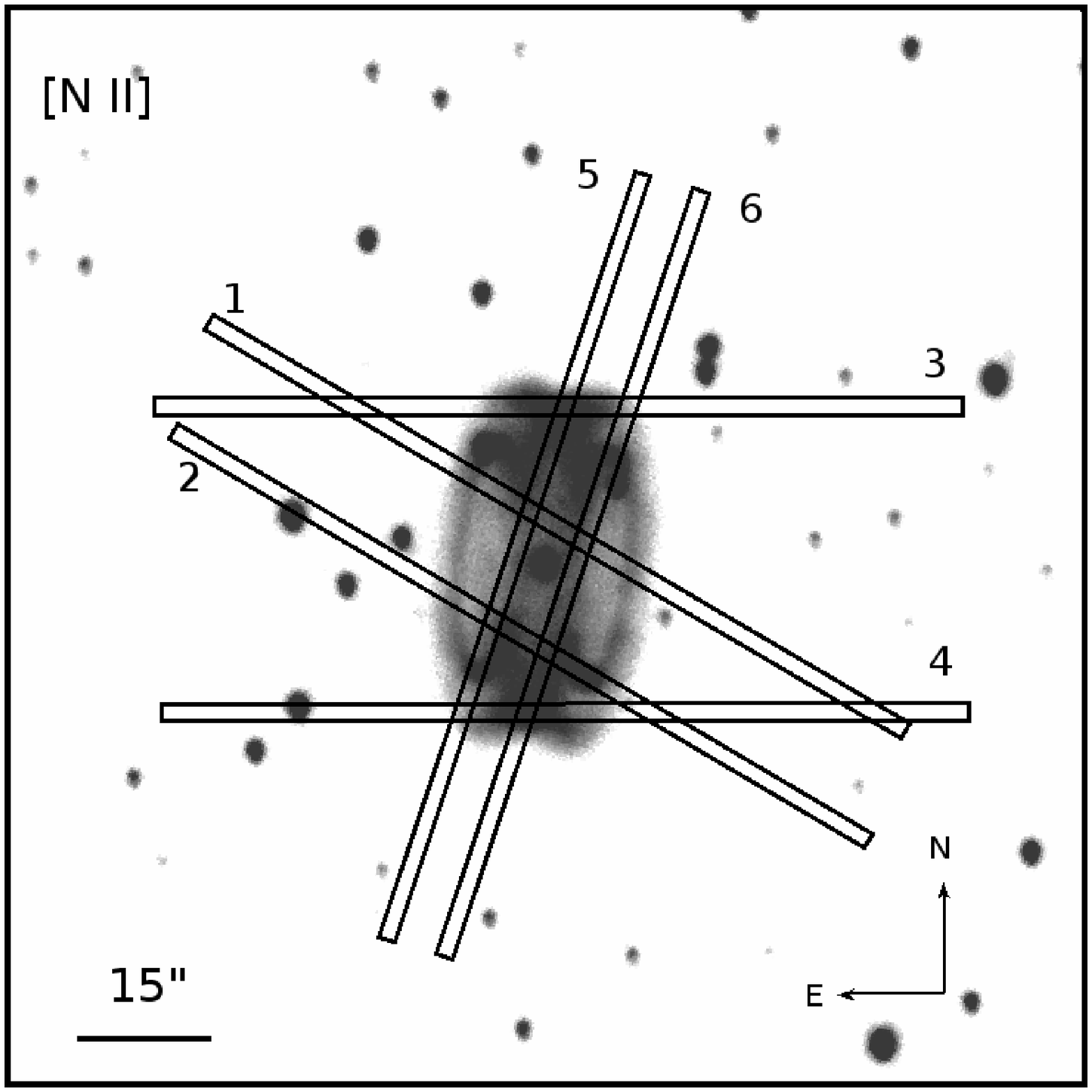}
\includegraphics[scale=0.28]{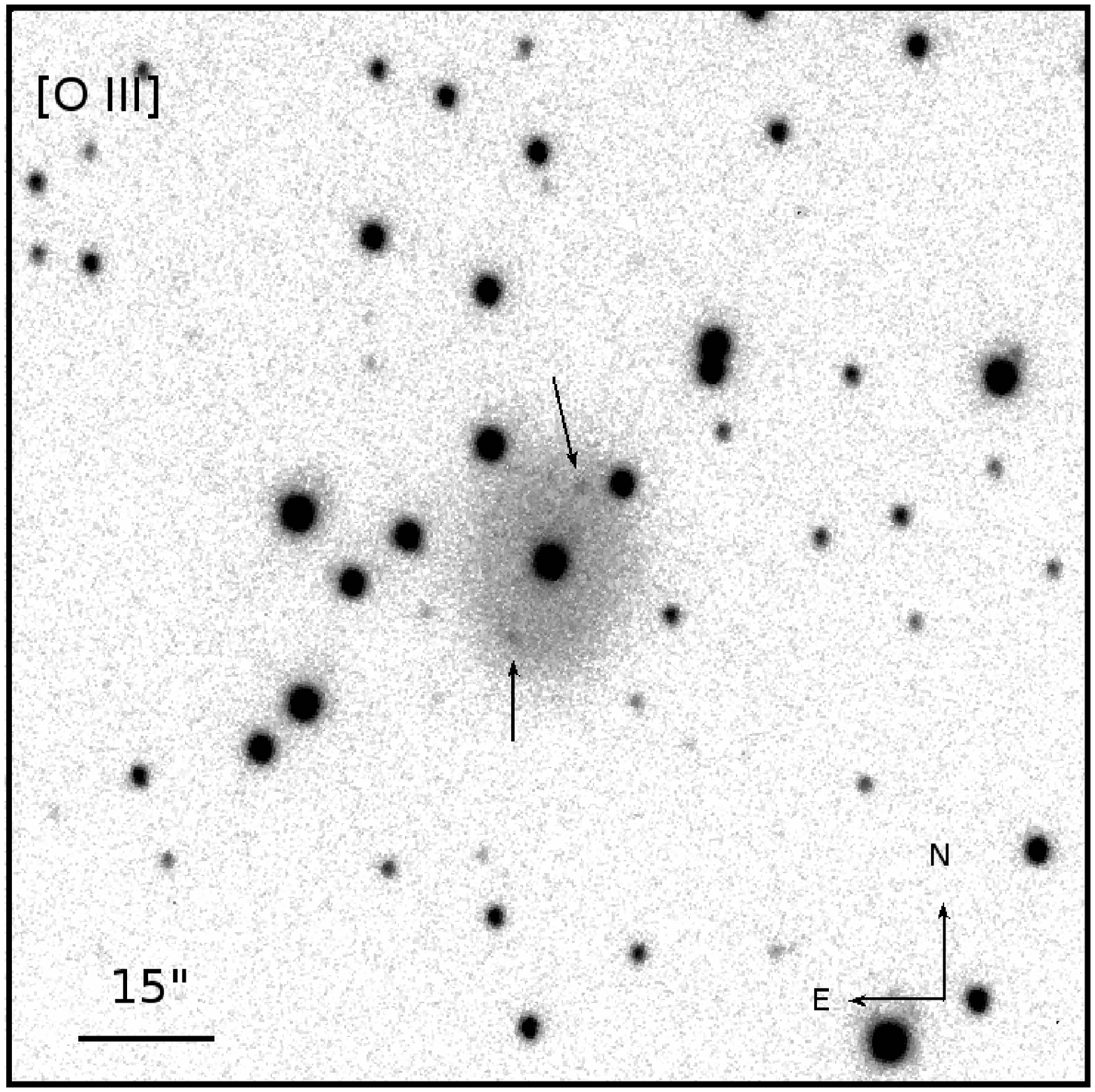}
\caption[]{\ha, \nitrogen, and \oxygeniii\ images of Abell~14 obtained with the 2.3~m Aristarchos telescope. 
The field size for the images shown is $120\times120$~arcsec. The MES--SPM slit positions are overlaid on the \nitrogen\ image. 
The width of each rectangle corresponds to a slit width of 1.9~arcsec and the slit length is equal to 90~arcsec.
The two arrows on the \oxygeniii\ image indicates the presence of two field stars.}
\label{fig1}
\end{figure*}

The formation mechanism for the equatorial density enhancement in the AGB wind is, however, still poorly understood.
The most likely mechanisms are (i) the presence of magnetic fields (Chevalier \& Luo 1994; Garc\'ia--Segura \& L\'opez 2000; 
Frank \& Blackman 2004; Vlemmings 2012) or (ii) the interaction of a close binary nucleus during the post--common 
envelope phase (Iben \& Livio 1993; Soker \& Livio 1994; Soker \& Rappaport 2001; Jones et al. 2012). 
The magnetic fields alone may not be unable to provide sufficient angular momentum needed to form aspherical PNe (Soker 2006, 
Nordhaus \& Blackman 2007, Garc\'ia--Segura et al. 2014). On other hand, only a small number of PNe have been found to host a close 
binary nucleus ($\sim$50; see Jones et al. 2015 and references therein), which conflicts with the number of known 
aspherical PNe (de Marco 2009, Miszalski et al. 2011a,b). Moreover, the number of $\sim$50 PNe with close 
binary nucleus corresponds to a small percentage of the total number of known PNe in our Galaxy ($\sim$3500; e.g. Kwitter et al. 2014; 
Sabin et al. 2014. In order to address this question, bipolar planetary nebulae that exhibit extreme or unusual 
properties can be investigated, and if these sources can be understood they will help in resolving the question of the 
shaping mechanism for PNe. 

The motivation for this work comes from the noticeable differences between the optical spectra (emission line intensities) 
and the chemical abundances of the PN Abell~14 or PN~A66~14 (Abell 1966) derived by Bohigas 2003 (hereafter B03) and Henry et al. 2010 
(hereafter H10), as well as its classification as a possible born--again nebula by de Marco (2009). 
We performed a morpho--kinematical study of Abell~14 by means of high--dispersion, long--slit 
spectra and the the astronomical code \textsc{shape}\footnote{http://www.astrosen.unam.mx/shape/} (Steffen \& L\'opez 2006, Steffen et al. 2011) 
in order to construct the 3D structure of the nebula. The 3D density distribution 
from our best--fit \textsc{shape} model was used as an input to the 3D photo--ionization code \textsc{mocassin} 
(Ercolano et al. 2003, 2005, 2008) in order to perform a more detailed study of its ionization structure and derive the fundamental 
physical parameters of the central source and the nebula.

The manuscript is organized as follows: In Section~2, we present a detailed description of Abell~14. The observations 
and data reduction procedures are presented in Section~3. In Section~4, we present the results from the 3D analysis of Abell~14, 
and describe our 3D modelling with the codes \textsc{shape} and \textsc{mocassin}. 
Our morpho--kinematic and physico--chemical results are discussed in Section~5. Conclusions are summarized in Section~6.

\section{Abell~14}

The central star of Abell~14 has been classified as a massive blue giant B5 III--V ($V$\,=\,15.24 and $B-V$\,=\,0.51) by Lutz \& Kaler (1987) and more recently as a B8--9 star by Weidmann \& Gamen (2011). Interestingly, these observational results do not agree with the stellar properties derived by studying the properties of its surrounding nebula (B03). In particular, the author has found that the central star of Abell~14 should be a hot star with a Zanstra temperature of 150,000~K and a low visual magnitude of $V_{\rm o}$\,=\,22.9\,$\pm\,0.3$, using the formula given in Jacoby \& Kaler (1989) for optically thick PNe, despite Abell~14 is an optically thin PN. This suggests that the central star of Abell~14 may be a binary system containing a hot, faint companion and a type B giant star. According to the evolutionary models from Vassiliadis \& Wood (1994) and Bl\"ocker (1995), B03 estimated the initial mass of the progenitor star to be between 2.5 and 3~$M_{\odot}$, with an evolutionary age of $\sim$8,000~yr.

B03 and H10 studied the chemical properties of Abell~14, and there are notable differences between the two studies. 
In particular, the intensities of the \oxygeniii\ $\lambda\lambda4959, 5007$ emission lines from 
H10 are $\sim$1.6 times larger than those from B03. The \oxygeni\ $\lambda6300$ line is detected by 
H10 with \oxygeni/\hbeta\,=\,30, while it is not found by B03 despite the detection of the fainter 
\helium\ $\lambda5876$ line (\helium/\hbeta\,=\,15.6). The \nitrogena\ $\lambda 5200$ line, which could be associated with strong shock 
interactions, is detected by B03 but not by H10. The logarithmic dereddened \hbeta\ fluxes are also slightly 
different between the two studies: $\rm{log}F$(\hbeta)\,=\,$-14.23$ (B03) and $-$14.82 (H10). These differences may be 
related to the slit positions where the spectra were obtained: along the north--south direction (PA=0\degree; B03), and 
along the east--west direction (PA=100\degree; H10). More recently, Frew et al. (2013) published a total integrated 
\ha\ flux of $-12.45$, which is close to the value of $-$12.27 quoted by B03.

Regarding the chemical abundances, H10 calculated higher He, N, and O than those derived in B03 
by factors of 1.25, 2.7, and 1.95, respectively, and lower S and Ar abundances by factors of 0.34 and 0.68, respectively. 
However, both studies agree that Abell~14 is an extremely nitrogen--rich nebula with $\rm{log}(N/O) = 0.48$ (B03) and 
0.62 (H10). A similarly high N/O abundance ratio has been found in highly evolved bipolar PNe such as PN~G321.6+02.2 (Corradi et al. 1997), 
HaTr~10 (Tajitsu et al. 1999), RCW~24 and RCW~69 (Frew et al. 2006), and PN~G342.0-01.7 (Ali et al. 2015). Each of these PNe exhibits very 
low electron density ($N_{\rm e}$) and very hot, optically faint central star, like that of Abell~14. The connection of these PNe with 
Abell~14 is discussed in \S~6.

\begin{table}
\centering
\caption[]{Observation log}
\label{table5}
\begin{tabular}{llllllllllll}
\hline 
\hline
Slit  & Filter     &  Date      & PA/Offset           & Exp.\\   
      &            &            & (\degree)/(\arcsec)  & (s) \\
\hline
& & 2.1~m SPM & Observatory & &\\
& & Spectroscopy            & &\\ 
\hline 
1     &  \ha+\nitrogen\      & 21--11--2012 & 59 / 4 N & 1800\\
2     &  \ha+\nitrogen\      & 21--11--2012 & 59 / 12 S & 1800\\
3     &  \ha+\nitrogen\      & 21--11--2012 & 87 / 17 N & 1800 \\
4     &  \ha+\nitrogen\      & 21--11--2012 & 87 / 17 S & 1800 \\
5     &  \ha+\nitrogen\      & 22--11--2012 & 341 / 5 E & 1800 \\
6     &  \ha+\nitrogen\      & 22--11--2012 & 341 / 5 W & 1800\\
6     &  \oxygeniii\         & 26--03--2015 & 341 / 5 W & 1800\\
\hline
& & 2.3~m Helmos & Observatory & & \\
& & Imaging                    & & \\
\hline
-   &  \ha\          & 08--11--2013 & $-$ & 1800\\
-   &  \nitrogen\    & 08--11--2013 & $-$ & 1800\\
-   &  \oxygeniii\   & 08--11--2013 & $-$ & 1800\\
\hline
\hline
\end{tabular}
\end{table}

Unlike the chemical abundances and line intensities, the $N_{\rm e}$\sulfurt, $T_{\rm e}$\nitrogen, and $c$(\hbeta) parameters are 
in agreement within the errors between the two aforementioned studies. The more recent value of $c$(\hbeta)\,=\,1.52, 
reported by Giammanco et al. (2011), is significantly higher than the previous reported values from B03 (0.98) and H10 (0.88). 
The extremely low $N_{\rm e}$ values of 55~$\rm{cm^{-3}}$ (H10) and 85~$\rm{cm^{-3}}$ (B03) indicate an old nebula. 
It should be mentioned that these values suffer from high levels of uncertainty. In particular, the errors of $N_{\rm e}$ have been estimated 
to be 165~$\rm{cm^{-3}}$ (H10) and $^{+70}_{-60}$~$\rm{cm^{-3}}$ (B03), respectively. According to this, an $N_{\rm e}$ value 
of $\le$200~$\rm{cm^{-3}}$ should be considered as more reasonable for this nebula.

Regarding the distance of Abell~14, it still remains poorly constrained in the literature with reported values of 3.32~kpc (Cahn et al. 1992), 3.38~kpc (Stanghellini \& Haywood 2010), 5.40~kpc (Giammanco et al. 2011), 7.25~kpc (Zhang 1995), and 2.66 to 3.77~kpc (Phillips 2004).
The bipolar structure of this nebula introduces a high level of uncertainty to its distance estimation and none of the aforementioned 
methods can be considered more reliable for this specific case. An average distance of 4.3~kpc with a standard deviation of 1.7~kpc 
is obtained from these measurements. This distance is found to be close to the value derived in this work based on 3D photo-ionization models (see \S 4.3).

\section{Observations}

\subsection{Optical imaging}

New high--quality optical images of Abell~14 were obtained with the 2.3 m Ritchey--Chretien 
Aristarchos telescope (f/8) at the Helmos Observatory in Greece on 2013 November 8. 
The observations were taken with a 1024\,$\times$\,1024 SITe CCD detector consisting of 24~$\upmu$m$^2$ pixels. 
The field of view and image scale were 5 $\times$\,5$~\rm{arcmin^2}$ and 0.28~arcsec~pixel$^{-1}$, respectively. 
Exposures of 1800~s were obtained through 17, 17, and 30~\AA~bandwidth filters centred on the \ha\ ($\lambda 6567$), \nitrogen\ ($\lambda 6588$), 
and \oxygeniii\ ($\lambda 5011$) nebular emission lines (Table~1). During observations, the seeing varied between 1.1 and 1.3~arcsec. 
Individual images were bias subtracted, flat--field corrected, and cleaned of cosmic rays using standard routines in \textsc{iraf}\footnote 
{\textsc{iraf} (Image Reduction and Analysis Facility) is distributed by the National Optical Astronomy Observatory, 
which is operated by he Association of Universities for Research in Astronomy (AURA) Inc., under cooperative agreement 
with the National Science Foundation.}. The processed images are shown in Fig.~1.

\begin{figure}
\centering
\includegraphics[scale=0.639]{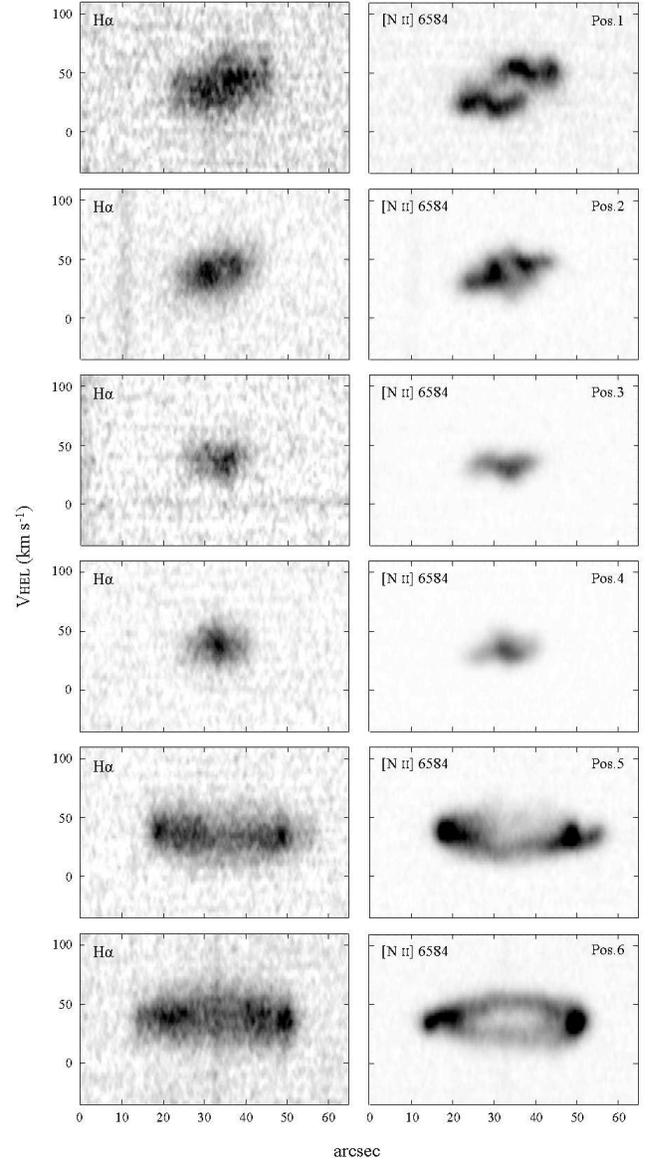}
\caption[]{\ha\ (left--hand column) and \nitrogen\ $\lambda 6584$ (right--hand column) PV diagrams of Abell~14 for 6 slit positions.}
\label{fig2}
\end{figure}

\subsection{High-resolution spectroscopy}

\begin{figure*}
\centering
\includegraphics[scale=0.63]{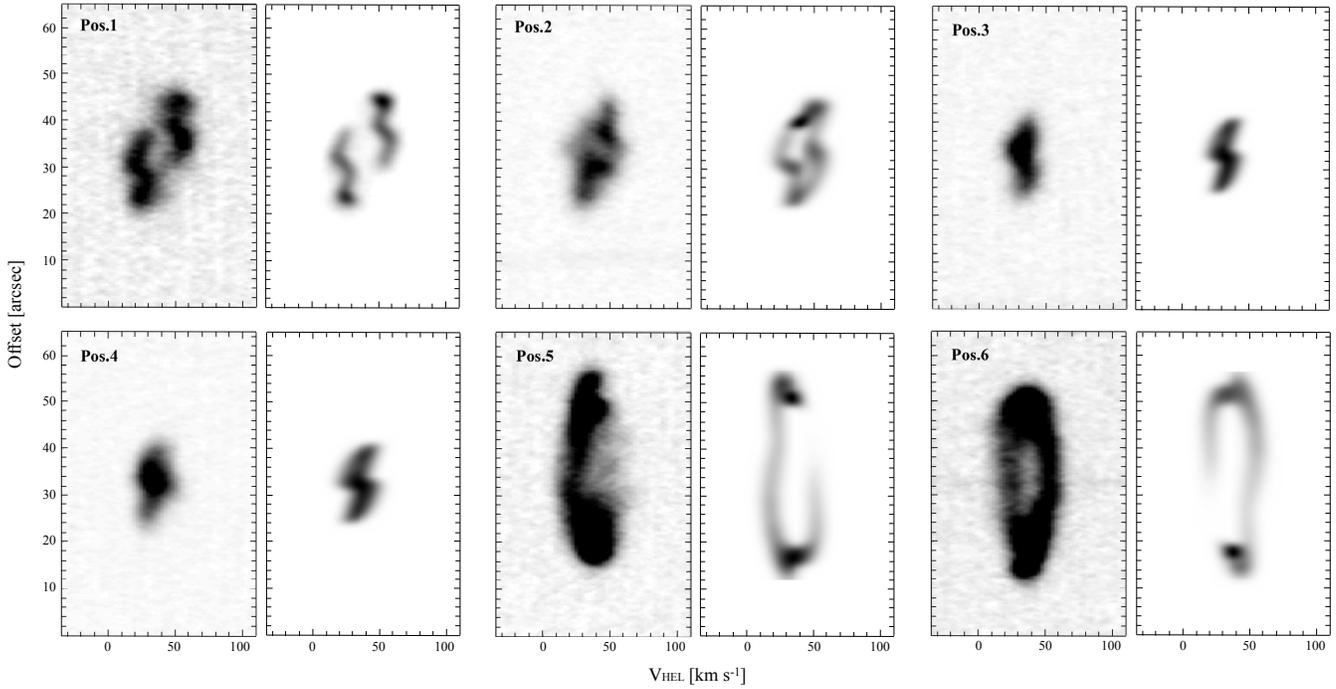}
\caption[]{Observed \nitrogen\ $\lambda 6584$ and synthetic PV diagrams of Abell~14 for 6 slit positions. The systemic velocity $V_{\rm sys}=40$~\kms.}
\label{fig3}
\end{figure*} 

High--resolution, long--slit spectroscopic data of Abell~14 were obtained in \ha+\nitrogen\ and \oxygeniii\ using the 
Manchester Echelle Spectrometer (MES--SPM; Meaburn et al. 2003) on the 2.1 m telescope at the San Pedro Martir Observatory 
in Baja California, Mexico, in its {\it f}/7.5 configuration. The observing runs took place on 2012, 21-22 November and 2015, 26 March (Table 1).

The MES--SPM was equipped with a Marconi 2048\,$\times$\,2048 CCD with each square pixel 13.5~$\upmu$m 
($\equiv$0.176~arcsec~pixel$^{-1}$) on each side. Bandwidth filters of 90 and 60~\AA\ were used to isolate the 87$^{\rm th}$ and 
113$^{\rm th}$ orders containing the \ha+\nitrogen{} $\lambda\lambda$6548, 6584  and \oxygeniii\ $\lambda 5007$ nebular emission lines. 
The \oxygeniii\ $\lambda 5007$ echelle spectrum is not presented here since it displays a simple ellispoidal structure 
with a very low expansion velocity. A binning of $2\times2$ was employed in both the spatial and spectral directions. 
Consequently, 1024 increments, each one 0.352~arcsec long, gave a projected slit length of $\sim$6~arcmin on the sky. The slit used was 
150 $\upmu$m{} wide ($\equiv$11.5~km $\rm{s^{-1}}$ or 1.9~arcsec). During the observations, the slits were oriented at different 
PAs (see Table 1) with an integration time of 1800~s for each orientation. All slit positions are shown overlaid on the \nitrogen\ 
image of Fig. 1.

The wavelength calibration was performed using a Th/Ar calibration lamp to an accuracy of $\pm$1~km~$\rm{s^{-1}}$ when 
converted to a radial velocity. The data reduction was performed using the standard \textsc{iraf} and 
{\textsc{starlink}\footnote{http://starlink.eao.hawaii.edu/starlink}} packages. Individual images were bias subtracted 
and cosmic rays removed. The bi--dimensional line profiles or position-velocity (PV) diagrams are presented in Fig. 2.

\section{Analysis and Results}

\subsection{Morpho--kinematic analysis}

Abell~14 displays an axially symmetric morphology in the \ha\ and \nitrogen\ emission lines with a size of 
$41\times23$~arcsec (Fig. 1). Two ring--like structures, each with a radius of 11.5~arcsec, 
are apparent to the east and west regions of the nebula, expanding with a de--projected velocity $V=17\,\pm$\,4~\kms. 
Their bright \nitrogen\ emission seen in Fig.~1 is likely associated with shock interaction. In the \oxygeniii\ image, 
the nebula shows more diffuse emission concentrated mostly in the central region with a size of $26\times15$~arcsec. 
This is consistent with the simple appearance of the \oxygeniii\ line in the PV diagram discussed above. Two spots 
on opposite directions of the central star, indicated by arrows on the \oxygeniii\ image of Fig.~1, are barely visible, and by 
scrutinizing the digital sky survey (DSS) images, we concluded that these two spots most likely correspond to very faint field stars.
 
The seemingly faint ring--like structure at the centre of the nebula, seen in \ha\ and \nitrogen\ lines, is the result of the projection effect, 
similar to those observed in two other PNe, namely SuWt~2 (Exter et al. 2010, Jones et al. 2010) and WeBo~1 (Tyndall et al. 2013). 
The nebular inclination and position angles with respect to the plane of the sky are found to be $\sim$24 and $\sim$5\degree, respectively.
 
Fig.~2 displays the observed PV diagrams for the \ha\ $\lambda 6563$ and \nitrogen\ $\lambda 6584$ emission lines for the slit 
positions 1 to 6. The PV diagrams in \ha\ reveal a more diffuse emission filling the internal volume of the nebula, 
whereas the \nitrogen\ emission originates from the outer parts of the nebula, like the walls of the bipolar lobes. 
The eastern and western parts of the nebula are blue-- and red--shifted respectively, both with a de--projected velocity of 
$V_{\rm{exp}}=25\pm4$~$\kms$. The heliocentric systemic velocity of the nebula is estimated to be $V_{\rm{sys}}=40\pm4$~$\kms$.

\subsection{\textsc{Shape} modelling}

A detailed morpho--kinematic study of Abell~14 was performed using the astronomical code \textsc{shape}. Modelling with \textsc{shape} mainly 
involves three steps. First, defining the geometrical forms to use; \textsc{shape} has a variety of objects such as sphere, torus, cone, cube, etc., 
whose basic forms can be modified by the user (e.g. squeeze, twist, bump, etc). Second, an emissivity distribution is 
assigned to each individual structure that makes up the nebula. A third, a velocity law is chosen as a function of 
position from the geometrical centre. \textsc{shape} produces a 2D image and synthetic PV diagrams that are rendered from the 3D model 
and compared visually with the observed data. The parameters of the model are then interactively adjusted until a satisfactory solution is obtained 
(e.g. Akras and Steffen 2012; Akras and L\'opez 2012; Clyne et al 2014; Clyne et al. 2015).

\begin{figure}
\centering
\includegraphics[scale=0.32]{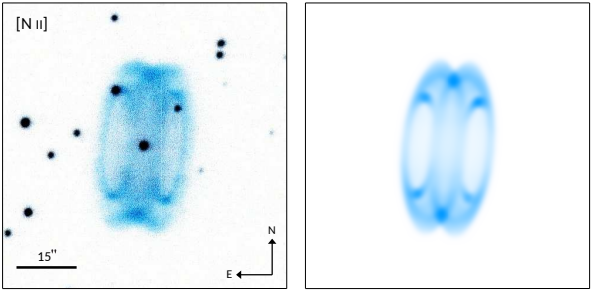}
\caption[]{\nitrogen\ $\lambda 6584$ image (left-hand) obtained from the Aristarchos telescope and the 2D rendered \textsc{shape} image (right-hand). 
The F.O.V. of each image is $75\times75$~arcsec.}
\label{fig4}
\end{figure}

Our model of Abell~14 was constructed using the image and kinematic data from the \nitrogen\ $\lambda 6584$ line due to its lower 
thermal broadening compared to that of the \ha~line (thermal width of the \ha~and \nitrogen~lines at 10,000~K are 21.4\kms~and 5.8\kms, 
respectively). This choise results in higher quality PV diagrams. An ellipsoidal shell was used to model the main body of the nebula, and was 
modified to conform with the observed image and PV diagrams. The model also includes two symmetrical rings that are positioned east and west of the nebular 
centre. The density of each component (ellipsoidal shell and both rings) was carefully modified by visually comparing and matching 
the contrast between the observed and synthetic PV diagrams. A homologous expansion velocity law of $v(r)=k(r/r_0$)\kms~was used, 
where $r$ is the distance from the geometrical centre and $r_{0}$ is the reference radius as measured from the 
\nitrogen\ image. A key characteristic of a cylindrically symmetric and homologously expanding nebula is how the PV diagrams can 
be altered in such a way that the outline of the image and the PV diagram match up. This matching determines the factor of 
proportionality $k$ in the mapping between position and velocity. It must however be noted that it is not possible to constrain 
the extent of the nebula along the line of sight and thus more than one solution can be found to fit the data.     

For our best--fit model, $k$=21\kms~arcsec$^{-1}$ and $r_{0}$=21.5~arcsec were derived for the main nebular structure. For each slit position, 
the resultant synthetic PV diagrams are placed side by side with the observed PV diagrams, as shown in Fig. 3. The observed \nitrogen\ image and 
that from the best--fit model are illustrated in Fig. 4. The 3D mesh structure of the model with the Doppler--colour velocity fields 
at six different orientations are presented in Fig. 5 for a better visualization of the 3D structure of the nebula.

The modelled inclination angle along the east--west direction with respect to the plane of the sky is $22^{\circ}\pm4^{\circ}$, 
whereas it is zero along the north--south direction. The de--projected expansion velocity of the outer \nitrogen\ shell in the equatorial 
direction is 21\kms. This implies that for a distance $D$ in kpc, and a constant expansion law, the kinematical age of the nebula is $tD^{-1}=4,850\,\pm$\,870
~yr~kpc$^{-1}$. Using the lowest (3.32~kpc) and highest (7.25~kpc) literature distances of Abell~14 (see \S 2), its equatorial size is 
estimated between 0.34 and 0.75~pc, which are reasonable values for highly evolved PNe (hereafter HE--PNe), whereas its kinematical age 
is between 16,000 and 35,000~yr. Notice that these ages do not agree with the evolutionary age of $\sim$8,000~yr, proposed by B03, because of the assumption that its central star has a $L\sim200$~$L_\odot$ which is several times larger than the value found in this work (see \S 4.3). 

\begin{figure}
\centering
\includegraphics[scale=0.46]{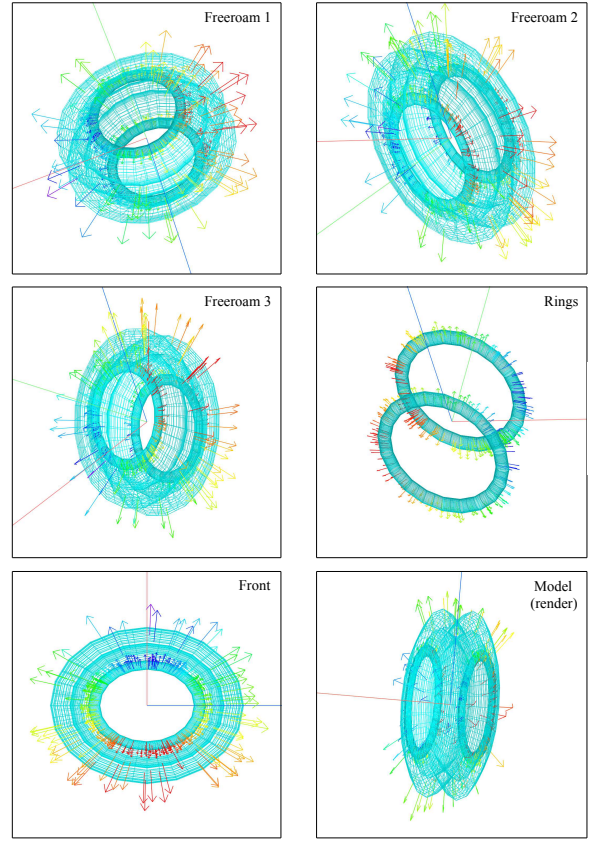}
\caption[]{Mesh images of Abell~14 in three random orientations, labeled as Freeroam 1, 2 and 3, showing its nebular components and velocity fields. 
The blue and red vectors correspond to the blue-- and red--shifted parts of each component.}
\label{fig5}
\end{figure} 

\begin{table*}
\centering
\caption[]{Observational fluxes and empirical abundance, as well as input parameters and line fluxes for the models studied in this work.}
\label{table5}
\begin{tabular}{lllllll}
\hline 
\hline
                     &       B03$^a$  &H10$^b$& MOD1$^c$ &  MOD2$^d$& MOD3$^e$& MOD4$^f$     \\           
\hline      
{\bf Source}          &                &      &          &          &          &                  \\
H:He:CNO($\%$)        &           $-$  &  $-$ & 91:9:0.1 & MOD1     & MOD1     & 0:33:67 \\
T$\star$ (kK)         &           151  &  $-$ & 150      & 150      & 120      & 120     \\
L$\star$/L$_{\odot}$  &          $-$   &  $-$ & 200      & 15       & 15       & 15      \\
$\rm{log}\rm{(g)}$    &           $-$  &  $-$ &  7       & 7        & 7        & 7       \\
\\
{\bf Abundance}       &                &      &          &          &          &         \\
He/H                  &           0.16 & 0.196&  0.16    &  0.196   & 0.16     & 0.16    \\       
C/H (10$^{-4}$)$^g$   &           $-$  & $-$  &  $-$     &  $-$     & 4.6      & 4.6     \\   
N/H (10$^{-4}$)       &          3.90  & 10.5 &  3.90    & 10.5     & 6.0      & 8.0     \\    
O/H (10$^{-4}$)       &          1.30  & 2.54 &  1.30    & 2.54     & 2.1      & 3.0     \\
Ne/H (10$^{-5}$)      &          $-$   & 9.79 &  $-$     & 9.79     & 9.8      & 9.8     \\    
S/H (10$^{-6}$)       &          9.30  & 3.15 &  9.30    & 3.15     & 12.0     & 20.0    \\
Ar/H (10$^{-6}$)      &          2.10  & 1.46 &  2.10    & 1.46     & 1.5      & 1.5     \\
Cl/H (10$^{-8}$)      &          $-$   & $-$  &  $-$     & $-$      & 1.0      & 1.0     \\
\\
\helium\ 5876         &          0.16  & 0.22 & 0.02     & 0.24     & 0.20     & 0.24    \\ 
\heliumb\ 4846        &          0.34  & 0.26 & 1.61     & 0.45     & 0.33     & 0.34    \\
\nitrogena\ 5200      &          0.29  & $-$  & 0.00     & 0.00     & 0.00     & 0.00    \\
\nitrogen\ 5755       &          0.29  & 0.41 & 0.02     & 0.16     & 0.16     & 0.05    \\
\nitrogen\ 6548       &          4.79  & 5.5  & 0.36     & 5.41     & 4.8      & 4.51    \\ 
\nitrogen\ 6584       &         14.35  &  16  & 1.09     & 16.2     & 14.3     & 13.6    \\
\oxygeni\  6300       &          $-$   &  0.3 & 0.00     & 0.02     & 0.06     & 0.02    \\ 
\oxygeniii\ 4363      &          $-$   &  $-$ & 0.10     & 0.05     & 0.04     & 0.01    \\
\oxygeniii\ 4959      &          0.98  &  1.6 & 1.71     & 1.72     & 1.40     & 0.75    \\ 
\oxygeniii\ 5007      &          3.07  &  4.8 & 5.14     & 5.15     & 4.20     & 2.24    \\
\sulfurt\ 6717        &          0.57  & 0.66 & 0.03     & 0.08     & 0.70     & 0.42   \\
\sulfurt\ 6731        &          0.43  & 0.49 & 0.02     & 0.06     & 0.53     & 0.32   \\
$-\rm{log}$(\hbeta)   &         13.24  & 13.85& 13.70    & 13.70    & 13.70    & 13.60  \\    
\hline 
{\bf Diagnostics}     &                 &      &         &          &          &         \\                     
$\rm{N_e}$:                       &     &      &         &          &          &         \\                      
$[$S~{\sc ii}$]\frac{6731}{6717}$ & 0.75& 0.74 & 0.79    & 0.76     & 0.76     & 0.76       \\                
$\rm{T_e}$:                       &      &      &         &          &          &           \\     
$[$N~{\sc ii}$]\frac{6584}{5755}$ & 65.9& 52   & 40.     & 134.     & 121.     & 355.    \\                        
$[$O~{\sc iii}$]\frac{5007}{4363}$& --- & ---  & 66.     & 130.     & 138.     & 260.    \\                
\hline
\hline
\end{tabular}
\begin{flushleft}
$^a$ Slit width 3~arcsec oriented at north--south direction (PA=0\degree).\\
$^b$ Slit width 2~arcsec oriented at PA=100\degree.\\
$^c$ Model using B03 abundances and slit configuration.\\
$^d$ Model using H10 abundances and slit configuration.\\
$^e$ Best--fit model using H10 slit configuration as constraints.\\
$^f$ Best--fit model using H10 slit configuration as constraints and PG~1159 central star.\\ 
$^g$ The solar carbon abundance (Asplund et al. 2009) is assumed for the models MOD3 and MOD4
\end{flushleft}

\end{table*}

For the ring--like structures, which have a radius of 11.5~arcsec and a de--projected velocity of 16\kms, we get a kinematical age of 
$tD^{-1}=3,400\pm620$~yr~kpc$^{-1}$ or between 11,300 and 24,700~yr for the low and high literature estimates of the distance to Abell~14. 
This implies that the ring--like structures were formed during a more recent ejection event. 

\subsection{\textsc{Mocassin} modelling}

In order to study the ionization structure of a bipolar PN, or even one with more complex 3D morphologies such as Abell~14, 
the use of sophisticated 3D photo--ionization codes is required. Our photo-ionization model was created with \textsc{mocassin} (version 2.02.70), 
which is described in full detail in Ercolano et al. (2003, 2005, 2008) and that uses the atomic data files from the CHIANTI database (v.~7, Landi et al. 2012). The procedure of modelling Abell~14 
requires to generate a density distribution of the nebular plasma and define the luminosity and effective temperature, the abundances for the most relevant 
elements and the distance of the nebula. Numerous models were generated until a good fit to the predefined constraints was found.
  
It is, however, a time consuming and trick task to generate an accurate 3D density distribution of the nebular plasma, which is why the density grid from our 
3D \textsc{shape} model was used as an input to the \textsc{mocassin} code. It is important to point out that our \textsc{shape} model was obtained 
using only the \nitrogen\ line, since the \oxygeniii\ line exhibits a simple ellipsoidal structure. In order to get a density grid for the entire nebula (inner and outer regions), an ellipsoidal structure with an uniform density was simply added to the \textsc{shape} density grid. This additional component is consistent with the observed \oxygeniii\ 2D image and spectrum. The last necessary step before modelling was to adjust the \textsc{shape} density grid, which comes in arbitrary units, in such a way that it successfully reproduces the \sulfurt\ $\lambda\lambda 6716, 6731$ line ratio as well as constraints (\hbeta\ flux, 2D-image, etc). The final structure used to produce the best modelled results is shown in Fig. 6.

Two sets of two long-slit spectra in different position angles obtained by B03 and H10 were used to constrain the model. The slit positions 
are shown in Fig.~7 overlaid on the best-fit model \oxygeni\ $\lambda 6300$ image. As these data were not ideal for performing a 
3D photo--ionization model -- spatially resolved data (e.g. integral field unit; IFU) would be really advantageous for reproducing the spatial emission 
distribution of the nebula and study the ionization structure -- it is important to keep this fact in mind when evaluating the model results 
presented in the next sections.

Besides the density distribution, another key element in photo--ionization modelling is the ionizing source. Recent developments in the modelling of stellar 
atmospheres have provided more complex stellar spectra that can be explored. Physical and chemical parameters such as $\rm{log}(g)$ and element 
abundances come into play. The standard choice of ionizing source models is the grid provided by T. Rauch, and described in detail in Rauch (2003), 
where the user can find NLTE stellar model atmosphere fluxes that cover the parameter range of PNe with hot central stars: ${T_{\rm eff}}=50-190$~kK, 
$\rm{log(g)}=5-9$, and distinct abundances. In this work, we explored the ionizing sources of the second generation models including elements H$-$Ni, 
detailed Ca, as well as the iron-group (Sc+Ti+V+Cr+Mn+Fe+Co+Ni) opacities (MOD3). We also explored fits with atmosphere grid calculated from models 
with typical PG 1159 star abundance ratios; He:C:N:O = 33:50:2:15 by mass (MOD4).

\begin{figure}
\centering
\includegraphics[scale=0.52]{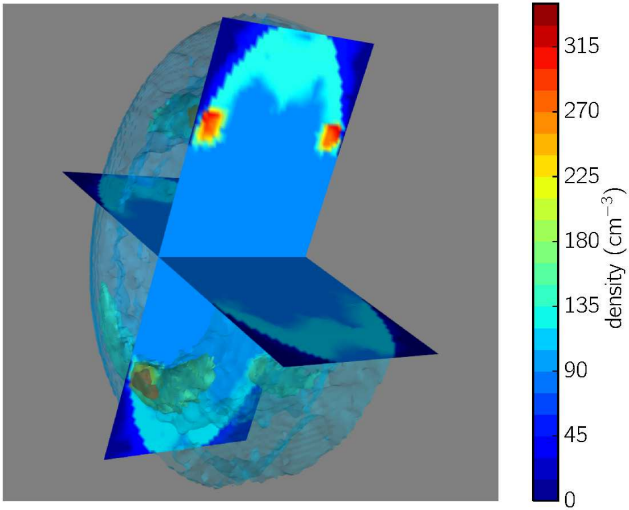}
\caption[]{Cuts through the structure grid used as the input density distribution for the \textsc{mocassin} model of Abell 14.}
\label{fig6}
\end{figure} 

Moreover, given that there are chemical abundances derived from the observed spectra (B03 and H10), we also ran models 
using the observed chemical abundances together with our density structure and ionizing source mentioned previously (see columns 2 and 
3 in Table~2). We find that the modelled spectra (MOD1 and MOD2) are dissimilar from the observed spectra. This may indicate the 
chemical abundances in Abell~14,derived by B03 and H10 using ionization correction factors (ICFs), are uncertain.
Gon\c calves et al. (2014) pointed out that an additional correction to the ICFs may be necessary when dealing with non-spherical PNe, 
such as Abell~14. Our results are summarized in Table~2. 

\begin{figure}
\centering
\includegraphics[scale=0.66]{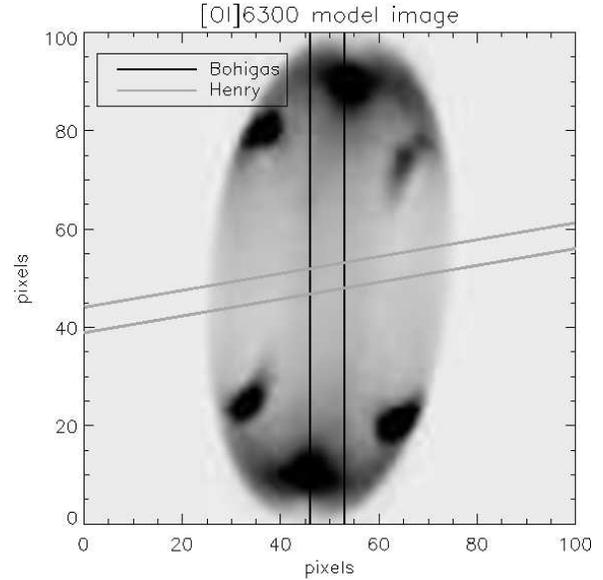}
\caption[]{Projected \oxygeni\ $\lambda 6300$ image obtained from the best--fit model with simulated slits overlaid. 
North is up and east is to the left.}
\label{fig7}
\end{figure}

Given the observational constraints and the substantial difference between the observed fluxes from B03 and H10, MOD3 reproduces H10's observed 
spectrum with an acceptable degree of matching. The mismatch of some emission lines shown may be directly associated with the presence of 
shocks in this nebula, which are not taken into account in our pure photo-ionization model. The typical error of the emission lines is of the order of 15\%. The contours from the projected model image (MOD3) are shown overlaid on the observed \nitrogen\ $\lambda 6584$ image in Fig. 8, where the model and observational images appear to be in good agreement. Our best-fit model was obtained for a distance of 4~kpc, whilst the central star has the following atmospheric parameters: ${T_{\rm eff}}=120$~kK, $\rm{log}\rm{(g)}=7$, and ${L/L_{\odot}}=15$ with an uncertainty of 10-15\%. This suggests that the central source of Abell~14 is a highly evolved star near the white dwarf (WD) cooling track with an evolutionary age around 10$^5$~yr, according to the evolutionary track from Vassiliadis and Wood (1994) and Bl\"ocker (1995). The former age is much larger than the kinematical age of the nebula of 19,400~yr for a distance of 4~kpc (the best fitting value based on our photo-ionization models). This discrepancy is commonly found in HE--PNe with a WD nucleus such as ESO~166--PN~21 (Pe\~na et al. 1997) and PN~S~174 (Tweedy and Napiwotzki 1994). One means by which the age discrepancy may be resolved is if the central source is a close binary system. Such a system would evolve more quickly than a single ionizing central source (Iben and Tutukov 1996). In the case of Abell~14, the observational data are likely consistent with a binary system scenario, but it has not yet been confirmed.

The same model also predicts ${T_{\rm e}}$\nitrogen=9,300~K, whereas the observed values from B03 and H10 are 12,000~K and 10,500~K, 
respectively. None of our models were able to reproduce these high ${T_{\rm e}}$, and probably an additional heating mechanism like 
shocks may have been taken place. The same scenario has also been proposed by B03. The 2D map of the modelled ${T_{\rm e}}$ is presented in Fig.~8 (lower left-hand panel), where it shows the stratification of ${T_{\rm e}}$ in the nebula. The inner parts are actually hotter than the outer parts by 3000-4000~K. The error of the modelled ${T_{\rm e}}$ is between 10-20\%, whereas the discrepancy between the model and the observations is between 15-30\%. This further supports the scenario of an additional heating mechanism. Besides the ${T_{\rm e}}$ discrepancy, the hypothesis of a shock mechanism in this nebula is also evident by the detection of the two neutral lines \nitrogena\ $\lambda 5200$ and \oxygeni\ $\lambda 6300$ emission lines (see Table~2). Our photo-ionization models could not  reproduce the intensities of these lines. Nevertheless, bright \nitrogena\ $\lambda 5200$ and \oxygeni\ $\lambda 6300$ lines have also been detected in the HE--PN G342.0--01.7 (Ali et al. 2015). These authors argue that a reverse shock into the nebula, due to the interaction of the nebula with the interstellar medium (ISM), results in the enhancement of the neutral lines. This could also be the case for Abell~14. 

\begin{figure}
\centering
\includegraphics[scale=0.40]{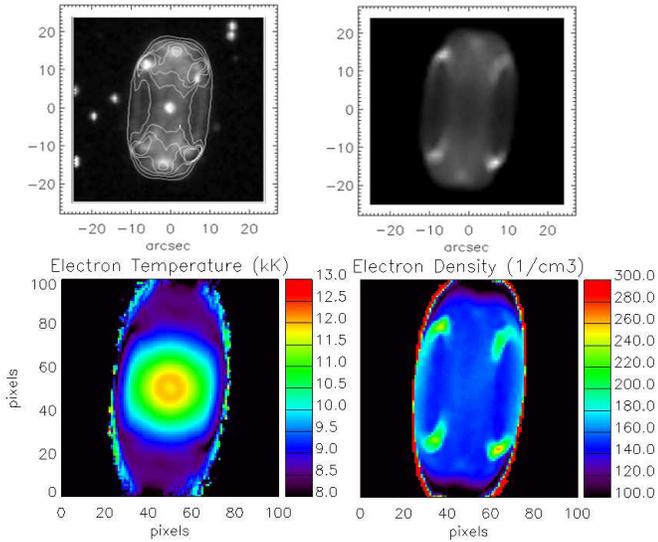}
\caption[]{Upper panels: Projected nitrogen\ $\lambda 6584$ image obtained from the best--fit model (left) and an observed narrow--band image (right). 
The white contours are from the model \nitrogen\ projected image. Lower panels: Modelled temperature and electron density maps. North is up and east is to the left.}
\label{fig8}
\end{figure}

We also attempt to reproduce a model assuming a hydrogen-deficient PG~1159 stellar atmospheric model with abundance ratios of He:C:N:O = 33:50:2:15 
by mass (MOD4). This model does not provide a good match with the observed spectra, whilst it predicts a very low  ${T_{\rm e}}$=6,600~K. 
We therefore consider that a hydrogen-deficient PG 1159-type star is likely not to be the case for the central source of Abell 14.

\begin{table*}
\centering
\caption[]{Chemical abundances$^{\dagger}$ and dereddened emission lines of HE--PNe}
\label{table5}
\begin{tabular}{llllllllllll}
\hline
\hline 
PN Name    & He       & N         & O     & N/O       & \nitrogen/\ha & \sulfurt/\ha & \nitrogena $\lambda 5200^a$ & \oxygeni $\lambda 6300^a$         & Reference\\   
\hline 
RCW 24     & $>$10.96 & 8.47      & 8.03: & 2.75     &   6.1          &  0.51         &         $-$                  &         52                       &Frew et al. 2006\\
RCW 69     &11.46:    & 8.70      & 8.37  & 2.14     &   6.9          &  0.65         &         29                   &         32                       &Frew et al. 2006 \\
ESO166--PN21$^b$& 11.14&$>$7.75$^b$& 8.60  & 0.61$^d$  &  1.7           &  0.27         &        $-$                   &         $-$                       &Pe\~na et al. 1997\\
G342.0--01.7&11.20     & 8.55      & 8.54  & 1.01     &  4.15          &  0.60         &        24                    &         55                       &Ali et al. 2015\\
HaTr 10    &   $-$    & 8.70      & 8.00  & 5.01     &  6.33          &  0.53         &        $-$                   &         $-$                     &Tajitsu et al. 1999\\ 
SBW~1      &   $-$    &$>$7.20$^c$& $-$   & $-$      &   0.6          &  0.13         &        $-$                   &         $-$                      &Smith et al. 2007\\
SBW~2      &   $-$    &$>$8.13$^c$& $-$   & $-$      &   7.42         &  0.58         &        $-$                    &        86.6                     &Smith et al. 2007\\
SuWt~2     &   $-$    &$>$8.40$^c$& $-$   & $-$      &   8.21         &  0.81         &        $-$                    &        59.2                     &Smith et al. 2007 \\
SuWt~2$^{\dagger\dagger}$&10.92 &8.60& 8.6   & 0.82     &   4.70         &  0.41         &        $-$                    &     119$^e$                 &Danehkar et al. 2013\\ 
WeBo~1     &   $-$    &$>$8.11$^c$& $-$   & $-$      &   2.63         &  0.09         &        $-$                    &        41.6                     &Smith et al. 2007 \\
G321.6+02.2&  11.17   & 9.22      & 8.75  & 2.95     &  12.27         &  0.79         &        16.24$^f$              &        30.57$^f$                &Corradi et al. 1997\\
Abell~14   & 11.20    & 8.59      & 8.11  & 3.02     &  6.69          &  0.35         &         29                    &       $-$                      &B03\\     
Abell~14   & 11.29    & 9.02      & 8.41  & 4.07     &  7.52          &  0.40         &        $-$                    &       30                       &H10\\
Abell~14$^{\dagger\dagger}$& 11.20&8.78& 8.32& 2.88     &  6.67          &  0.43         &        $-$                    &       $-$                      &this work (MOD3)\\     
Abell~14$^{\dagger\dagger}$& 11.20&8.90& 8.48& 2.63     &  6.34          &  0.26         &        $-$                    &       $-$                      &this work (MOD4)\\     
\hline
Type--I     &11.14     & 8.77      & 8.76  & 1.03     &  Fig.~9       & Fig.~9         &        $-$                    &       $-$                      &Henry et al. 2004\\
Type--II    &11.05     & 8.20      & 8.70  & 0.31     &  Fig.~9       & Fig.~9         &        $-$                    &       $-$                      &Henry et al. 2004\\
The Sun     &10.93     & 7.83      & 8.69  & 0.14     &  $-$          & $-$            &        $-$                    &       $-$                      &Asplund et al. 2009\\
\hline
\hline
\end{tabular}
\medskip{}
\begin{flushleft}
$^{\dagger}$ Chemical abundances are given in the notation of 12+$\rm{log}$[n(X)/n(H)], $^{\dagger\dagger}$ These values have been derived from photo--ionization models.\\
$^a$ Relative to \hbeta~= 100, $^b$ Average value from 7 slit positions, $^c$ This value is a lower limit and corresponds only to N$^+$.\\
$^d$ This ratio was calculated based on the assumption of N/O = N$^+$/O$^+$, $^e$ The night sky emission line at 6300\AA\ may have not been properly subtracted.
$^f$ They have been dereddened using R = 3.1
\end{flushleft}
\end{table*}

Also of importance is the significant discrepancy in the chemical abundances between the studies by B03 and H10. Our best--fit model predicts an 
N abundance value between both studies, an O abundance close to H10's value, and an S abundance close to that of B03 (see Table~2). The chemical abundances of 
Abell~14 remain poorly constrained, meriting further study. Despite the significant difference between the chemical abundances, each study agrees 
that there is a significant N enrichment in the nebula. The N/O ratio is found to be substantially high with values of 3 (B03), 4.13 (H10), 2.85 (MOD3), and 2.66 (MOD4). 
These values, in conjunction with the high He abundance [0.16 (B03) and 0.196 (H10)], clearly indicate that the progenitor star has experienced the TDU phase. 
Given that the O abundance is between 2 (B03) and 3 (H10) times lower than the solar value (Asplund et al. 2009), large amounts of O and C must have been 
converted to N via the CN-- and ON--burning cycle (HBB), indicating an intermediate--mass progenitor of $M\geq3~M_{\odot}$ 
(Marigo 2007). According to the single star evolutionary tracks of Vassiliadis and Wood (1994), only a star with an initial mass of $\sim 5~M_{\odot}$, and a sub--solar metallicity (Z = 0.08), would be hot ($T_{\rm eff}=120$~kK) and faint enough ($L=15$~$L_\odot$) to match our findings after $\sim$85,000~yr. 
A higher stellar luminosity would result in a younger central star that would be consistent with the kinematical age of the nebula.

\section{Discussion}

This work presents the first detailed study of the peculiar PN Abell~14 using new high-dispersion echelle spectra and high-resolution images, as well 
as the astronomical codes \textsc{shape} and \textsc{mocassin}. The former code was used to create the 3D morphology and key features seen in the 2D 
images and PV diagrams, whereas the latter was used to reproduce its 3D ionization structure, fed by the density distribution obtained with the former. 

\subsection{Morpho--kinematic structure}

Abell~14 has a hollow bipolar shape with the lobes protruding out in the east--west directions. The \nitrogen\ emission originates mainly from the 
walls of the bipolar lobes, whereas the \oxygeniii\ emission originates from the central region. Two ring--like structures at the eastern and western 
parts of the nebula are apparent in the \ha\ and \nitrogen\ lines, but are absent in the \oxygeniii\ line.

The main bipolar shell expands with a velocity that increases with distance from the geometric centre.
Adopting an expansion velocity of 21~\kms\ for the equatorial region, the kinematical age of the nebula is calculated to be $tD^{-1} = 4,850\pm870$~yr~kpc$^{-1}$. A second event likely formed the two ring--like structures that are interacting with the main bipolar component. The kinematical age of the rings is $tD^{-1}=3,400\pm620$~yr~kpc$^{-1}$. Taking into account the lowest (3.32~kpc) and highest (7.25~kpc) available distance measurements of Abell~14 (see \S 2), we estimate the kinematical ages for the nebula and the rings to range from 16,000 to 35,000 and 11,300 to 24,700~yr, respectively. Adopting the distance of 4~kpc derived from our photoionisation modelling, the kinematical age is 19,400$\pm$3,500~yr. We consider this age more reliable.

The morphology of Abell 14 strongly resembles those of other ring-like PNe such as WeBo~1 (Tyndall et al. 2013) and SuWt~2 (Exter et al. 2010, Jones et al. 
2010), observed from different orientations. A comprehensive study involving several circumstellar rings has also been performed by Smith et al. (2007). 
These authors conclude that ring--like structures are divided into two groups: i) those that are formed around massive supergiants, like SN~1987A 
(Meaburn et al. 1995, Panagia et al. 1996) and ii) those around intermediate--mass stars in close binaries, like SBW~1 (Smith et al. 2007). 
The bipolar structure of Abell~14 is more consistent with the second scenario, although the binary system has not been yet confirmed as in the cases of 
WeBo~1 and SuWt~2.

In particular, WeBo~1 is a member of a small group of PNe (Abell~35, Abell~70, LoTr~1, and LoTr~5) whose central stars share several common properties (Bond et al. 1993, 2003). 
Their central stars are binary systems with a rapidly rotating giant or sub--giant and a hot, optically faint companion. Siegel et al. (2012) 
photometrically confirmed the binary nature of WeBo~1. Lutz \& Kaler (1987) and Weidmann \& Gamen (2011) proposed the presence of a binary system in Abell~14 
in order to explain the observational data. Morphologically, WeBo~1 clearly shows a ring--like structure in the \ha+\nitrogen\ emission lines, whereas the \oxygeniii\ emission 
is more diffuse in the inner region, like what we observe in Abell~14 (see Fig. 11 from Tyndall et al. 2013).

As for SuWt~2, it also exhibits a bright ring--like structure with much fainter bipolar lobes, seen almost edge--on (Exter et al. 2010, Jones et al. 2010). 
Its morphology is similar to that found in WeBo~1 and Abell~14, whilst its central star has been classified as a B9 or early A--type star (Exter et al. 2010). 
Such a cool star emits insufficient amounts of ionizing radiation, which suggests the presence of a hotter binary companion, as recently confirmed by Exter 
et al. (2010). Given the noticeable similarities among Abell~14, WeBo~1, and SuWt~2, they may be similar objects.

\subsection{Ionization structure and the central star}

\begin{table*}
\centering
\caption[]{Stellar and physical parameters of HE--PNe.}
\label{table5}
\begin{tabular}{llllllllllll}
\hline 
\hline
PN Name    &  ${T_{\rm eff}}$ (kK) & $L$ (${L_{\odot}}$) & $\rm{log}\rm{(g)}$ & ${N_{\rm e}}$ (cm$^{-3}$) & $D$ (kpc) & Kinematical age (yr)  & Reference\\   
\hline 
RCW~24     &  $\sim$120       & 70         &  $-$                & $<$100    & 1        &$>$ 26,000$^a$  &Frew et al. 2006\\
RCW~69     &  $\sim$130       & 190        &  $-$                & 250       & 1.3      &$>$ 17,000$^a$  &Frew et al. 2006 \\
ESO~166--PN21 & 69.2          & 22         & 7.14                & $\sim$100 & 1.19     & 16,100         &Pe\~na et al. 1997\\
G342.0--01.7& 105             & 118        & 7                   & $<$20     & 2.06     & $\geq$ 20,000  &Ali et al. 2015\\
HaTr~10    & 100              & 47         & $-$                 & 580       & 1.5       &$>$ 17,000$^a$  &Tajitsu et al. 1999\\ 
SBW~1      & $-$              & $-$        & $-$                 & 513       & 7        & 9,600          &Smith et al. 2007\\
SBW~2      & $-$              & $-$        & $-$                 & 280       & 2.3      &$>$~26,000      &Smith et al. 2007\\
SuWt~2$^{{\dagger}b}$ &160       & 600        & 7.3              & $<$100    & 2.3      &23,400--26,300   &Danehkar et al. 2013 \\
WeBo~1     & $-$              & $-$        & $-$                 & 312       & 1.6      &11,700-12,000    &Tyndall et al. 2013, Bond et al. 2003 \\
G321.6+02.2& $>$130           & $-$        & $-$                 & 250       & 2        & $>$~12,000      &Corradi et al. 1997\\
Abell~14$^{\dagger}$&120        & 15         & 7                 & 100-180   & 4        &19,400          &this work\\     
\hline
\hline
\end{tabular}
\medskip{}
\begin{flushleft}
$^{\dagger}$ The stellar parameters have been derived from photo--ionization models.\\
$^a$ These kinematical ages have been calculated by us, assuming a nominal expansion velocity of 40~\kms.\\
$^b$ A possible born--again PN.\\
\end{flushleft}
\end{table*} 

The 3D density distribution grid, derived using \textsc{shape}, was used as an input to the \textsc{mocassin} code in order to perform 
a more comprehensive study of its ionization structure. Two available spectra from the literature were used to constrain the models. It is worth pointing out that this technique of deriving independent density grids has also been applied to NGC~1501 by Ercolano et al. (2004) using a density grid derived from long-slit echellograms as an input to MOCASSIN and also to NGC~40 by Monteiro and Falceta-Gon\c{c}alves (2011), using a density distribution derived from 2.5D hydrodynamic simulations.

Two models were run with the chemical abundances as free parameters and assuming a stellar model atmosphere (MOD3) and a hydrogen-deficient, 
PG~1159 stellar atmosphere model (MOD4) for the central source. Both models adequately reproduced the majority of the emission lines 
(Table~2) for a distance of 4~kpc. The stellar parameters ($L$ = $15~L_{\odot}$, $\rm{log}\rm{(g)}=7$, and ${T_{\rm eff}}=120$~kK) indicate a highly evolved central star near to the cooling track of WDs. In accordance with the theoretical hydrogen-burning evolutionary track of Vassiliadis and Wood (1994), the central star has an evolutionary age of $\sim$85,000~yr (for $M=5~M_{\odot}$ and Z = 0.08), which is several times higher than 
the kinematical age of the nebula ($\sim$19,400~yr). Nevertheless, this discrepancy is a common problem in evolved PNe with 
highly evolved central stars (e.g. Pe\~na et al. 1997). A possible solution is the interaction between two binary components that would
accelerate the evolution of the system (Iben and Tutukov 1986). Given that the star at the centre of the nebula has been classified as a B--type massive blue giant (B5 III--V, Lutz \& Kaler 1987; B8-9, Weidmann \& Gamen 2011), the presence of a binary system in Abell~14 is very possible.
The helium--burning evolutionary track (very late thermal pulse) gives for the aforementioned stellar parameters an extremely high 
evolutionary age, that is between 200,000--300,000~yr. This age is 3-4 times larger than the evolutionary age based on hydrogen-burning tracks
and at least 10 times larger than the kinematical age of the nebular gas. Notice that SuWt~2, a likely 
born-again nebula (Danehkar et al. 2013), is one of the oldest member in our sample of HE-PNe, based on its kinematical age, but at 
the same time it is the most luminous. This indicates that the central star of SuWt~2 is in an earlier evolutionary stage than all other members (see Table 4). According to this analysis, Abell~14's stellar parameters ($L$ = $15~L_{\odot}$, ${T_{\rm eff}}=120$~kK) and its kinematical age ($\sim$19,400~yr)
cannot be explained by the born-again scenario, which should result in a much older nebula, for the given stellar parameters.

Our two best--fit models and the observed spectra agree that there is significant He and N abundance enrichment in Abell~14. Based 
on the Kingsburgh and Barlow (1994) classification scheme, Abell~14 is an extreme Type--I highly evolved nebula and probably 
one with the largest He abundance and N/O ratio. These values are consistent with a 5~$M_{\odot}$ progenitor derived from the 
evolutionary tracks.

\subsection{Comparison of Abell~14 and other highly-evolved PNe}

Until now, only a small number of studies have been performed on highly evolved Type--I PNe (see Tables 3 \& 4). All PNe in this group show 
substantially high N/O ratios and He abundances  whilst their central stars have very low $L$ and high $T_{\rm eff}$, and they are 
possible descendants of intermediate-mass stars in close binaries. Only the central star of SuWt~2 has a high luminosity of $600~L_{\odot}$ and 
a large kinematical age (Danehkar et al. 2013), due to a likely very late thermal pulse (born--again scenario). 
Moreover, while most HE--PNe show N/O ratio larger than 1 (N rich), SuWt~2 has an N/O = 0.8 which makes it somewhat different. 

Most of the HE--PNe in our group (Tables 3 and 4) lie in the low-left corner of the PNe regime on the \ha/\nitrogen\ vs. \ha/\sulfurt\ diagnostic diagram 
(Sabbadin et al. 1977; Gon\c calves et al. 2003) of Galactic PNe (Fig.~9). Akras \& Gon\c calves (2016) have recently demonstrated that in this regime also lie shock--excited regions, e.g. low--ionization structures such as knots and jets. Since these PNe do not exhibit high velocities, the scenario of shocks in such large 
and HE--PNe seems unlikely. Interestingly, some HE--PNe exhibit uncommonly bright \nitrogena\ $\lambda 5200$ and \oxygeni\ $\lambda 6300$ emission lines, 
which are strong indicators of shock activity. Recently, Ali et al. (2015) showed that a reverse shock of only 25~\kms\ in the PN G342.0--01.7, 
due to its motion through the ISM, can successfully explain the strong neutral emission lines. The low photo--ionization rate of their central stars, 
(very low $L$) and large nebular size, would make the contribution of shocks easily perceptible even at small velocities (Akras \& Gon\c calves 2016), 
or produce emission line spectra that resemble those of shock-excited regions (Raga et al. 2008).

Three out of 11 HE--PNe are seen to exist outside the normal region in Fig.~9, namely SBW~1, ESO~166--PN~21, and WeBo~1. SBW~1 and WeBo~1 are the youngest members in this group, and their central stars are such luminous that their high photo-ionization rate overcomes the shock excitation. Unfortunately, the stellar parameters of these HE--PNe are still unknown. Regarding ESO~166--PN~21, the \nitrogen/\ha\ and \sulfurt/\ha\ line ratios correspond to an average value from 7 slit positions (for more details see in Pe\~na et al. 1997). Studying the spectrum of each slit position separately, we deduce that the most distant regions (A1 and A5 in Pe\~na et al. 1997) exhibit stronger \nitrogen\ and \sulfurt\ lines. Two out of 7 regions in ESO~166--PN~21 lie in the regime of HE--PNe. This is in agreement with our previous analysis that the lower the photo-ionization rate, the stronger the low--emission lines (Raga et al. 2008, Akras \& Gon\c calves 2016). 

\section{Conclusions}

\begin{figure}
\centering
\includegraphics[scale=0.26]{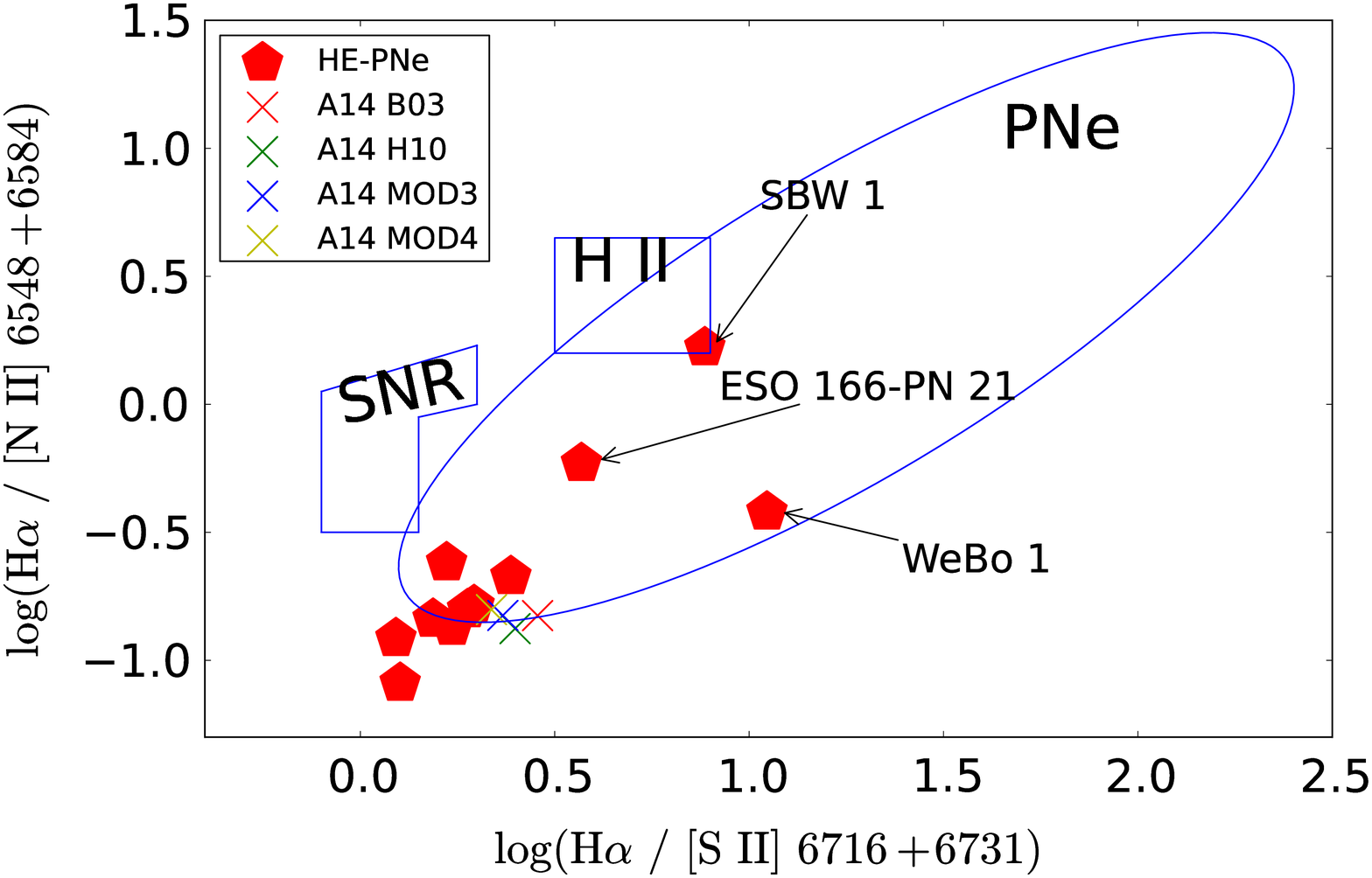}
\caption[]{A diagnostic diagram of Galactic PNe by Sabbadin et al. (1977) with the density ellipses introduced by Riesgo \& L\'opez (2006) and for the HE--PNe in Tables 3 and 4.}
\label{fig9}
\end{figure}

The first detailed study of the 3D morphology and ionization structure of Abell~14 has been discussed in this paper.
The kinematical and physical parameters of this nebula, as well as the stellar parameters of its central source have been derived using new 
high-dispersion spectroscopic data and high-resolution images, as well as the 3D codes \textsc{shape} and \textsc{mocassin}. 
The resultant modelled 3D density distribution obtained from \textsc{shape} has been used as an input to the \textsc{mocassin} code. 
To successfully create a photo-ionization model using the proposed technique (\textsc{shape + mocassin}), at least two emission lines 
(e.g. \nitrogen\ and \oxygeniii) are required to constrain the morphology of the inner and outer parts, while obtaining a 3D density 
distribution that best describes the entire nebula.

Abell~14 was found to be a highly evolved (19,400~yr), Type--I PN (11.20 $<$ He $<$ 11.29, and 2.63 $<$ N/O $<$ 4.03) formed by an intermediate-mass progenitor of 5~$M_{\odot}$, which it has likely experienced the TDU and HBB, resulting in a substantial enrichment of He and N.
The abundance of O is found to be sub--solar, indicating that large amounts have been converted to N. The synthetic PV diagrams and 2D image from the 
\textsc{shape} code at different orientations resemble the structure of WeBo~1 and SuWt~2, thus suggesting that these are objects with similar morphologies observed from different angles. 
The binary companion of the central star in WeBo~1 and SuWt~2 has already been confirmed, but not for the central source in Abell~14.
The bright star at the centre of Abell~14 has been classified as a massive blue-giant; but such star is not hot enough to ionize the surrounding gas, making a second, hotter companion probable. Our best--fit ionization models adequately reproduced the observed spectra for a central source with ${T_{\rm eff}}=120$~kK and $L=15\,L_{\odot}$. These stellar parameters can not be explained by the born-again scenario within an evolutionary timescale of $\sim$85,000~yr (hydrogen burning timescale). The central star of Abell~14 is in a more evolved evolutionary phase than that of SuWt~2, a likely born-again nebula. while it is surrounded by a younger nebula (lower kinematical age). This implies that the central star of Abell~14 has not experienced any very late thermal pulse event. 

Returning to the shaping mechanisms of bipolar planetary nebulae, this study indicates that a highly evolved, low luminosity and intermediate mass progenitor, within a binary system is responsible for the distinctive double-ringed structure and peculiar spectrum observed in Abell 14.

Abell~14 also shows unexpectedly strong \oxygeni\ $\lambda 6300$ and \nitrogena\ $\lambda 5200$ emission lines that are not reproducible by 
our photo--ionization models. This indicates that shock interactions may play an important role. The very low luminosity of its central star, in conjunction with the nebula's large size, imply a very low photo--ionization rate that makes the contribution of shocks easily perceptible. Regarding its \nitrogen/\ha\ and \sulfurt/\ha\ line ratios, Abell~14 seems to have some similarities with supernova remnants, but it is more closely related to other HE-PNe.

Of all HE--PNe discussed here, 8 out of 11 exhibit strong neutral \oxygeni\ $\lambda 6300$ line, likely associated with shocks, 6 out of 7 have
central stars with $L<200~L_{\odot}$, and are therefore at a very evolved stage near the WD cooling track with very low photo-ionization rates. 
We argue that any shock interaction due to stellar winds, or interaction between the nebula and the surrounding ISM (e.g. G342.0-01.7) may be the main reason for the enhancement of the low-ionization lines.

\section*{Acknowledgments} 

We also thank the anonymous referee for her/his thorough review and helpful comments.
The authors would like to thank the staff at SPM and Helmos Observatories 
for their excellent support during these observations. The Aristarchos telescope
operated on Helmos Observatory by the Institute of Astronomy, Astrophysics,
Space Applications and Remote Sensing of the National Observatory of Athens.
SA is supported by CAPES post--doctoral fellowship \lq Young Talents Attraction' --
Science Without Borders, A035/2013. NC gratefully acknowledges the support of 
the College of Science, NUI Galway (under their PhD fellowship scheme).
CHIANTI is a collaborative project involving George Mason University, 
the University of Michigan (USA) and the University of Cambridge (UK).

\bibliographystyle{mnras}

\end{document}